 \documentclass[preprint,review,12pt]{elsarticle}

\usepackage{amssymb}
\usepackage{graphicx}
\graphicspath{ {./} }
\usepackage{amsmath}
\usepackage{amssymb}
\usepackage{tikz}
\usepackage{secdot}
\usepackage{epstopdf}
\usepackage{geometry}
\usepackage{hyperref}
\usepackage{tabularx,ragged2e,booktabs,caption}
\usepackage{cite}
\usepackage{fixltx2e}
\biboptions{numbers,sort&compress}
\usepackage{float}

\journal{Journal of Physics and Chemistry of Solids}

\begin{document}
 
\begin{frontmatter}

\title{Phase components in Zr$_7$Ni$_{10}$ and Hf$_7$Ni$_{10}$ binary alloys; investigations by perturbed angular correlation spectroscopy and first principles calculations}

\author{S.K. Dey$^{1,2}$\corref{cor1}}
\ead{skumar.dey@saha.ac.in}
\author{C.C. Dey$^{1,2}$}
\cortext[cor1]{corresponding author}
\ead{chandicharan.dey@saha.ac.in}
\author{S. Saha$^{1,2}$}
\ead{satyajit.saha@saha.ac.in}
\author{G. Bhattacharjee$^{1,2}$}
\ead{gourab.bhattacharjee@saha.ac.in}
\author{D. Banerjee$^{3}$}
\ead{dbanerjee@vecc.gov.in}
\author{D. Toprek$^4$}
\ead{toprek@vin.bg.ac.rs}
\address{$^1$Saha Institute of Nuclear Physics, 1/AF Bidhannagar, Kolkata-700 064, India}
\address{$^2$ Homi Bhabha National Institute, Anushaktinagar, Mumbai-400 094, India}
\address{$^3$ Accelerator Chemistry Section, RCD(BARC), Variable Energy Cyclotron Centre, Kolkata-700 064, India}
\address{$^4$Institute of Nuclear Sciences Vinca, University of Belgrade, P. O. Box 522,\\ 11001 Belgrade, Serbia}

\begin{abstract}
Intermetallic
compounds Zr$_7$Ni$_{10}$ and Hf$_7$Ni$_{10}$ have been studied by perturbed
angular correlation (PAC) spectroscopy considering the fact that Zr$_7$Ni$_{10}$ has application as hydrogen storage material in fuel cell. In stoichiometric Zr$_7$Ni$_{10}$, the phase Zr$_2$Ni$_7$ is found to be dominant ($\sim$38\%) while a fraction of $\sim$25\% is found for the Zr$_7$Ni$_{10}$ phase at room temperature. In this compound, a phase due to Zr$_8$Ni$_{21}$ ($\sim$10\%) is found from room temperature up to 773 K but, this is not found at 873 K and above. In the stoichiometric Hf$_7$Ni$_{10}$ sample, the phase due to Hf$_7$Ni$_{10}$ is found as a minor phase ($\sim$22\%) at room temperature. In this system, no phase of Hf$_2$Ni$_7$ is observed but, a different phase due to HfNi$_3$ is found to be dominant ($\sim$62\%). It is found that the site fraction of Hf$_7$Ni$_{10}$ enhances with temperature at the expense of HfNi$_3$ and this phase becomes predominant ($\sim$57\%) at 673 K and above. The change of phase fractions  of HfNi$_3$ and Hf$_7$Ni$_{10}$ with temperature is found to be reversible. The phase components in Zr$_7$Ni$_{10}$
and Hf$_7$Ni$_{10}$ have been determined also from X-ray powder diffraction (XRD) and transmission
electron microscopy/selected area electron diffraction (TEM/SAED) measurements. Ab-initio calculations
using the all electron full potential (linearized) augmented plane
wave [FP-(L)APW] method, within the framework of 
the density functional theory (DFT), have been performed to determine the electric field gradients at the $^{181}$Ta impurity sites and therefrom to assign the different components observed from PAC measurements. Present experimental and calculated results of EFG support the $Cmca$ space group for both Zr$_7$Ni$_{10}$ and Hf$_7$Ni$_{10}$ compounds.
\end{abstract}

\end{frontmatter}

\section{Introduction}
The alloys based on Zr-Ni system have numerous technological applications. These are used as an integral part to form high
temperature eutectics \citep{Chandrasekhar}, bulk glassy alloys \citep{Yokoyama, Pang}, corrosion resistant material \citep{Zander,Pang},
shape memory alloys \citep{Matthew}, superalloys \citep{Singer, Ivanchenko}, superconductors \citep{Alzahrani}. Magnetic properties in the transition
metal based Zr$_9$Ni$_{11}$ alloy was found earlier \citep{ProvenzanoZr9Ni11}. Some of the Zr-Ni binary
alloys, viz. Zr$_8$Ni$_{21}$, Zr$_7$Ni$_{10}$, 
ZrNi and Zr$_9$Ni$_{11}$ \citep{Ruiz, Nei, Kwo, Young, Ruiz2010, Young2011, M Latroche} have received
considerable attention due to their ability to absorb large
amount of gaseous hydrogen and reversibility in hydrogen dissociation process. These alloys form metal hydrides (MH) after hydrogen absorption and these MHs
are used as negative electrodes in Ni-MH rechargeable batteries. The Ni-MH batteries have been widely used
in hybrid electric vehicles (HEV) due to their high energy density. Electrochemical capacity is found to be maximum
for Zr$_7$Ni$_{10}$ among Zr-Ni binary alloys \citep{Kwo}. Recently, it has
been found that the hydrogenation property of TiFe alloy is improved due to the addition of small fraction of Zr$_7$Ni$_{10}$ \citep{Jain}. Another recent study \citep{Amol} reported improvement in hydrogenation kinetics and capacity of Ti-V-Cr body centered cubic (BCC) solid solutions when Zr$_7$Ni$_{10}$ was used as an additive. The Hf-Ni alloys also
have many technological applications. The
alloys of Ni-Ti-Hf exhibit shape memory behavior \citep{Meng}. Intermetallic compounds of Hf and transition metals (Fe, Co, Pd, Pt) have also hydrogen 
storage properties \citep{Baudry}, with high H/M ratio at room temperature. 
 
Considering the above technological applications of these compounds, Zr$_7$Ni$_{10}$ and Hf$_7$Ni$_{10}$ binary alloys have been studied by
perturbed angular correlation (PAC) spectroscopy to determine the phase components produced in the stoichiometric samples and their stabilities with temperature. To the best of our knowledge, no previous investigation in Zr$_7$Ni$_{10}$ was done by PAC spectroscopy.
In Hf$_7$Ni$_{10}$, however, PAC measurements 
were carried out by Gil et al. \citep{Gil}. These authors found five electric quadrupole frequencies from PAC measurements. But, the components were not
assigned by comparing with DFT calculations.

  The crystal structure of Zr$_7$Ni$_{10}$ was first reported to be a non-centrosymmetric orthorhombic structure with space group $Aba2$
by Kirkpatrick et al. \citep{Kirkpatrick}. Later on, Joubert et al. \citep{J.-M. Joubert} corrected the crystal structure to be centrosymmetric 
orthorhombic structure with space group $Cmca$. The compound Hf$_7$Ni$_{10}$ is reported to be isostructural
with Zr$_7$Ni$_{10}$  \citep{J.-M. Joubert, Kirkpatrick, PAnash_Hf_Ni}. 

The perturbed angular correlation is an useful nuclear technique 
to study the structural properties, phase transitions, crystalline defects
and magnetic ordering in intermetallic compounds \citep{Schatz}. In this technique, the angular correlation of a suitable $\gamma$-$\gamma$ cascade of the probe nucleus
is perturbed by hyperfine interaction. The nuclear moments (electric quadrupole moment/magnetic dipole moment) of the probe nucleus interact
with hyperfine fields (electric field 
gradient/magnetic field) present at the probe site in the investigated sample. The electric field gradient (EFG) at the probe site is created by the surrounding 
charge distribution with a non-cubic symmetry and it is sensitive to the change of local electron density. Production of 
multiple phases in intermetallic alloys can be identified by this technique from the observation of different EFGs in the sample. Phase stability of compounds can be 
observed from temperature dependent PAC measurements. Internal magnetic field
of a ferromagnetic material
can be determined through the measurement of Larmor precession frequency. 
Using this technique, EFGs and magnetic ordering in Hf/Zr-Ni 
systems were studied earlier \citep{Marszalek2, Poynor, Koteski, Ivanovski, Marszalek, PRJSilva, Dey, Das, Srivastava}. Recently we have studied two Zr-Ni compounds which have hydrogen storage properties. These are Zr$_8$Ni$_{21}$ \citep{SKDeyZr8Ni21}, Zr$_9$Ni$_{11}$ \citep{SKDEYZr9Ni11}. In the present report, studies in another hydrogen absorbing material viz. Zr$_7$Ni$_{10}$ and its Hf analogue compound have been carried out by PAC, X-ray diffraction (XRD) and transmission
electron microscopy/selected area electron diffraction (TEM/SAED) techniques.

Present investigations have been carried out to determine the phases components in these alloys. Temperature dependent PAC measurements in the range 77-1073 K have been carried out to determine the phase stability. Assignment of phases and site preference of the probe atom have been done by comparing the measured values of EFG with the calculated results by density functional theory.

\section{Experimental details}
The intermetallic alloys Zr$_7$Ni$_{10}$ and Hf$_7$Ni$_{10}$ were prepared by arc-melting high purity Hf, Zr and Ni metals in stoichiometric amounts in an argon atmosphere.  
These metals were procured from Alfa Aesar. The purity of Zr was $\sim$ 99.2\% excluding Hf (maximum Hf concentration 4.5 wt\%) and the purity of 
Hf was $\sim$ 99.95\% excluding Zr (maximum Zr concentration 3 wt\%). The purity of Ni used was 99.98\%.
For preparation of Zr$_7$Ni$_{10}$, stoichiometric amounts of Zr and Ni were taken and melted homogeneously in the arc furnace by repeated melting. A shiny globule of Zr$_7$Ni$_{10}$ was formed which was then remelted with a tiny piece ($<$1 at\%) of natural Hf metal wire. The sample was then activated to $^{181}$Hf by irradiating with thermal neutrons using a flux of $\sim$10$^{13}$/cm$^2$/s at Dhruba reactor, Mumbai. For preparation of Hf$_7$Ni$_{10}$, a tiny Hf metal ($\sim$1 at\%) was first activated to $^{181}$Hf. The active Hf metal sample was then remelted with the sample Hf$_7$Ni$_{10}$, prepared in the arc furnace. The probe $^{181}$Hf resides at the Zr sites for 
Zr$_7$Ni$_{10}$ sample due 
to their chemical and structural similarity. In Hf$_7$Ni$_{10}$, the probe replaces identical Hf atoms.
These samples were then sealed in evacuated quartz tubes for PAC measurements
at high temperatures. Inactive samples 
of Zr$_7$Ni$_{10}$ and Hf$_7$Ni$_{10}$ were prepared separately in the arc furnace in a similar manner for XRD and TEM/SAED measurements. The XRD measurements 
were carried out by TTRAX-III x-ray diffractometer (Rigaku, Japan) using the Cu K$_\alpha$ radiation. Transmission electron
microscopy (TEM) measurements were carried out using
FEI, Tecnai G2 F30, S-Twin microscope equipped with a high angle annular dark-field (HAADF) detector, a
scanning unit and a energy dispersive X-ray spectroscopy (EDX) unit to
perform the scanning transmission electron
microscopy (STEM-HAADF-EDX).

In the PAC technique, the probe $^{181}$Hf undergoes a $\beta^-$ decay ($T_{1/2}$=42.4 d) to $^{181}$Ta and emits two successive $\gamma$-rays of energy 133 keV and 482 keV passing through an intermediate level (482 keV) with $T_{{1/2}}$=10.8 ns and
spin angular momentum $I=5/2^+\hbar$ \citep{Firestone}. The angular correlation
of 133-482 keV $\gamma$-$\gamma$ cascade is perturbed by the interaction
of electric quadrupole moment ($Q$=2.36 b \citep{Butz}) of the 
intermediate level and the surrounding EFG.

The perturbation function $G_2(t)$ for $I$ = 5/2 $\hbar$ in a polycrystalline sample is given by \citep{Schatz,Berkes,Zacate}, 
\begin{equation}
  G_2(t)=S_{20}(\eta) + \sum^{3}_{i=1}S_{2i}(\eta)\text{cos}(\omega_it)
  \text{exp}(-\delta\omega_it).
\label{eqn:Stokes}
 \end{equation}
 The frequencies $\omega_i$ are the transition frequencies between the three sublevels of the intermediate level arising due to hyperfine splitting. The $S_{2i}$ parameters depend on the asymmetry of the electric field gradient and these are expressed as a polynomial in asymmetry parameter ($\eta$). Due to lattice strain or defects present in a real crystal, the probe may get displaced from the actual lattice site. Different probes are thus subjected to slightly different electronic and ionic environment in the same phase of the crystal. This effect is considered by an exponential (Lorntzian) distribution function where $\delta$ is mean frequency distribution 
 width. A least squares fitting to 
  eqn. (\ref{eqn:Stokes}) 
  determines the quadrupole frequency ($\omega_Q$) through the observed transition frequencies $\omega_1$, $\omega_2$ and $\omega_3$. In the principal axis system, the EFG tensor has zero off-diagonal elements. Conventionally, the largest 
  component of the EFG tensor is designated as $V_{zz}$ which is related to quadrupole frequency by  
  \begin{equation}
\omega_Q= \frac{eQV_{zz}}{4I(2I-1)\hbar}.
 \label{eqn:raman}
\end{equation}
 The principal EFG components ($V_{xx}$, $V_{yy}$ and $V_{zz}$) obey the Laplace's equation
      \begin{equation}
V_{xx}+V_{yy}+V_{zz}=0, \quad \text{where}\quad   |V_{zz}|\ge |V_{yy}|\ge |V_{xx}|.
  \label{eqn:newton}
 \end{equation}
 The symmetry of the EFG is determined by the asymmetry parameter $\eta$, defined as 
 \begin{equation}
  \eta=\frac{V_{xx}-V_{yy}}{V_{zz}}, \quad \text{where} \quad  0\le \eta\le 1.
  \label{eqn:eta}
 \end{equation}
 Therefore, EFG can be determined from only two parameters $V_{zz}$ and $\eta$.
 For $\eta=0$, the perturbation function becomes periodic and harmonic, and the quadrupole frequency is related to the transition frequencies ($\omega_i$) by
 \begin{equation}
 \omega_Q=\omega_1/6=\omega_2/12=\omega_3/18.
 \label{eqn:prafulla}
\end{equation}
When $\eta\ne$0, the perturbation function remains periodic but not harmonic. The relations between $\omega_Q$ and $\omega_i$ become more complex and can be found in the reference \citep{Zacate}. 

Present PAC measurements have been carried out using a four detector BaF$_2$-BaF$_2$ or a four detector LaBr$_3$(Ce)-BaF$_2$ PAC set up. The crystal sizes were 38.1(dia) $\times$ 25.4(ht) mm$^2$ for LaBr$_3$(Ce) and 50.8(dia) $\times$ 50.8(ht) mm$^2$ for BaF$_2$. In the LaBr$_3$(Ce)-BaF$_2$ setup, the 133 keV $\gamma$-rays were selected in the LaBr$_3$(Ce) detector. Standard slow-fast coincidence assemblies were formed to collect data
at 180$^\circ$ and 90$^\circ$ \citep{pramana}. Typical prompt time resolutions (FWHM) of $\sim$1 ns and $\sim$790 ps were obtained at $^{181}$Hf energy window settings
for the BaF$_2$-BaF$_2$ and LaBr$_3$(Ce)-BaF$_2$ setup, respectively. Details of experimental set up and data analysis can be found in reference \citep{pramana}.
\begin{table}[t!]
	\begin{center}
\caption{Results of PAC measurements in Zr$_7$Ni$_{10}$}
\label{tab:table_Zr7Ni10} 
\scalebox{0.5}{
	\begin{tabular}{ l l l l l l l l l l l|} 
		\hline  \\ [-0.9ex]
		Temperature (K)   &Component     & $\omega_Q$ (Mrad/s)     & $\eta$     & $\delta$($\%$)   & $f$($\%$)      & Specification     \\ [1.5ex]
		\hline 
		77                & 1    & 77.9(3)                 & 0.36(2)          & 0.9(6)                & 34(3)                & Zr$_2$Ni$_7$      \\
		&2    & 63.8(3)                 & 0.52(1)          & 0               & 38(3)          &   Zr$_7$Ni$_{10}$  \\   
		&3    & 8.0(4)                 & 0               & 0                & 28(3)          &     \\ \hline   	
		298                &1    & 72.7(3)                 & 0.12(4)             & 1.1(7)                & 38(3)                 & Zr$_2$Ni$_7$      \\
		&2    & 58.9(3)                & 0.71(1)          & 0                & 25(3)         & Zr$_7$Ni$_{10}$      \\  
		&3    & 77.1(9)                 & 0.81(2)          & 0                & 11(3)          &  Zr$_8$Ni$_{21}$   \\   
		&4    & 8.0(7)                 & 0          & 0                & 17(3)             &  \\ 
		&5    & 33(1)                 & 0          & 0                & 9(3)             &  \\ \hline	
		373                &1    & 70.7(5)                & 0.23(5)        & 0                & 25(3)    &  Zr$_2$Ni$_7$   \\  
		&2    & 57.9(3)               & 0.70(2)         & 0                & 21(3)    &  Zr$_7$Ni$_{10}$         \\   
		&3    & 79(1)                 & 0.81(5)         & 0                & 13(3)   & Zr$_8$Ni$_{21}$              \\ 
		&4    & 6.7(3)                 & 0         & 0                & 28(3)     &            \\ 
		&5    & 32.7(8)                 & 0         & 0                & 13(3)     &            \\ \hline		
		473                &1    & 69.4(2)               & 0.20(3)          & 0                & 21(3)       & Zr$_2$Ni$_7$       \\  
		&2    & 56.6(2)                 & 0.74(1)          & 0                & 19(3)             &   Zr$_7$Ni$_{10}$ \\   
		&3    & 80(1)                 & 0.82(4)         & 0                & 10(3)          &  Zr$_8$Ni$_{21}$     \\ 
		&4   & 5.7(2)                 & 0          & 0                & 32(3)               & \\ 
		&5   & 32.3(5)                 & 0          & 0                & 17(3)               & \\  \hline                   
		573                &1    & 67.4(2)                 & 0.12(4)          & 0                & 34(3)              & Zr$_2$Ni$_7$ \\ 
		&2   & 54.4(3)                & 0.80(1)          & 0                & 26(3)       &  Zr$_7$Ni$_{10}$    \\    
		&3   & 80(1)                 & 0.82(4)         & 0                & 7(3)            &   Zr$_8$Ni$_{21}$    \\ 
		&4   & 7.7(5)                 & 0          & 0                & 24(3)               &  \\ 
		&5   & 34(1)                 & 0          & 0                & 9(3)               &  \\ \hline               
		673                &1    & 65.4(2)                & 0          & 0                & 31(3)     &   Zr$_2$Ni$_7$       \\   
		&2    & 52.5(5)                 & 0.85(2)          & 0                & 21(3)              & Zr$_7$Ni$_{10}$\\ 
		&3    & 77(1)                 & 0.87(4)        & 0                & 7(3)         &   Zr$_8$Ni$_{21}$      \\ 
		&4    & 5.9(3)                 & 0          & 0                & 30(3)           &      \\ 
		&5    & 32.9(9)                 & 0          & 0                & 11(3)           &      \\ \hline
		773                &1    & 63.0(2)                & 0          & 1.2(4)                & 29(3)     &Zr$_2$Ni$_7$     \\   
		&2    & 47.9(4)                 & 0.86(1)          & 0                & 19(3)              & Zr$_7$Ni$_{10}$ \\ 
		&3    & 77(1)                 & 0.96(7)          & 0                & 8(3)         &   Zr$_8$Ni$_{21}$      \\ 
		&4    & 5.3(2)                 & 0          & 0                & 29(3)           &      \\ 
		&5    & 31.7(5)                 & 0          & 0                & 15(3)           &      \\ \hline               
		873                &1    & 62.4(2)                & 0.18(4)          & 0                & 37(3)     & Zr$_2$Ni$_7$      \\   
		&2    & 44.7(9)                 & 0.93(9)          & 0                & 18(3)              & Zr$_7$Ni$_{10}$ \\ 
		&3    & 5.5(3)                 & 0          & 0                & 31(3)         &        \\ 
		&4    & 30(1)                 & 0          & 0                & 13(3)           &      \\ \hline
		973                &1    & 61.5(2)                & 0.12(6)          & 0                & 27(3)     & Zr$_2$Ni$_7$        \\   
		&2    & 44.7(7)                & 0.84(4)          & 0                & 24(3)              & Zr$_7$Ni$_{10}$  \\ 
		&3    & 4.0(2)                 & 0          & 0                & 31(3)           &      \\ 
		&4    & 28.4(8)                 & 0          & 0                & 17(3)           &      \\ \hline                   
		1073              &1    & 62.7(1)              & 0.14(3)          & 0                &31(3)     &  Zr$_2$Ni$_7$    \\   
		&2    & 46.0(5)                & 0.88(3)          & 0                & 18(3)              & Zr$_7$Ni$_{10}$  \\ 
		&3    & 4.6(1)                 & 0          & 0                & 35(3)           &      \\ 
		&4    & 29.9(5)                 & 0          & 0                & 15(3)           &      \\ \hline                                  
		298$^*$           &1     & 73.1(3)                 & 0.12(7)          & 0                & 35(3)   & Zr$_2$Ni$_7$       \\   
		&2    & 59.8(7)                 & 0.65(4)          & 0                & 31(3)             & Zr$_7$Ni$_{10}$   \\ 
		&3    & 74(1)                 & 0.79(5)          & 0                & 10(3)        &   Zr$_8$Ni$_{21}$      \\ 
		&4    & 7.3(6)                 & 0          & 0                & 17(3)             &    \\ 
		&5    & 32(2)                 & 0          & 0                & 6(3)             &    \\ 		\hline
	\end{tabular}}
	\begin{center}
		\scriptsize{$^*$ After measurement at 1073 K } 
	\end{center}
\end{center}
\end{table}	
	
	\begin{figure*}
		\begin{center}
			\includegraphics[width=0.8\textwidth]{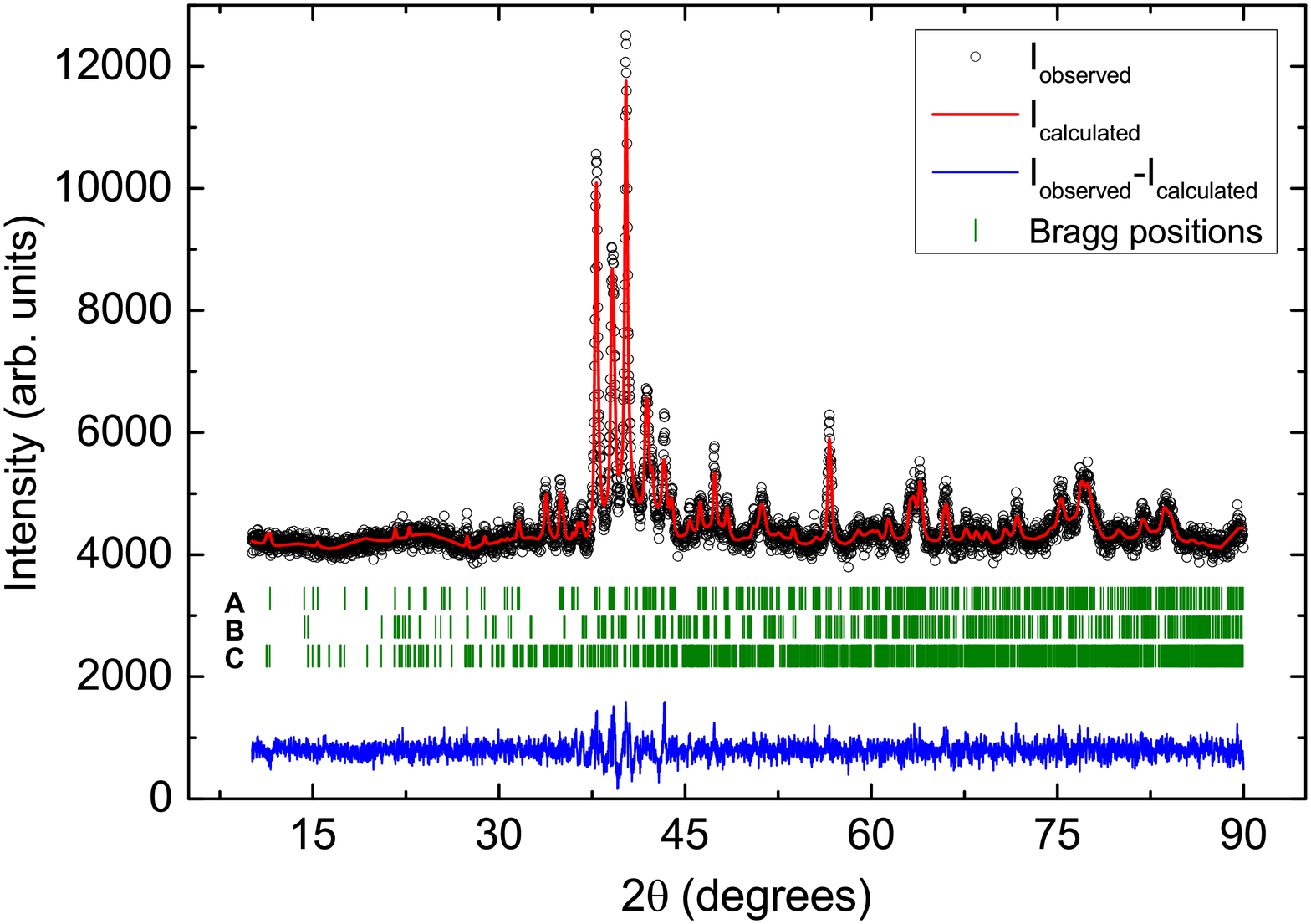}
		\end{center}
		\captionof{figure}{\small{The background subtracted XRD powder pattern in the stoichiometric sample of Zr$_7$Ni$_{10}$. The vertical bars A, B and C denote
				the Bragg angles corresponding to Zr$_7$Ni$_{10}$, Zr$_2$Ni$_7$ and Zr$_8$Ni$_{21}$ phases,
				respectively.}}
		\label{XRD_Zr7Ni10}
	\end{figure*} 
	
	\begin{figure}
		\begin{center}
			\includegraphics[width=0.45\textwidth]{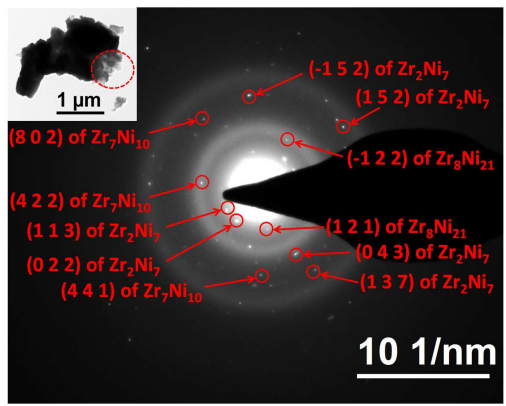}
		\end{center}
		\captionof{figure}{\small{Selected area electron diffraction pattern from stoichiometric
				Zr$_7$Ni$_{10}$ particle shown in the
				inset.}}
		\label{TEM_Zr7Ni10}
	\end{figure} 
	
	\begin{figure*}
		\begin{center}
			\includegraphics[width=0.6\textwidth]{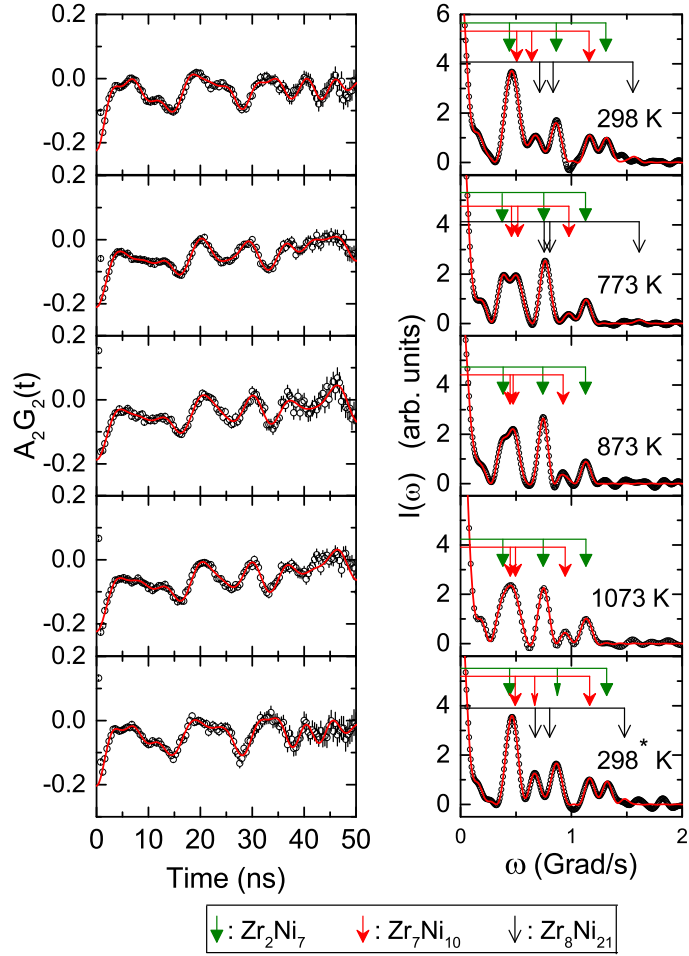}
		\end{center}
		\captionof{figure}{\small{Perturbed angular correlation spectra in Zr$_7$Ni$_{10}$ at different temperatures. 
				Left panel shows the time spectra and the right panel shows the corresponding Fourier cosine transforms. The PAC spectrum designated
				by 298$^*$ K is taken at room temperature after the measurement at 1073 K.}}
		\label{Zr7Ni10PAC}
	\end{figure*} 
	
	\begin{figure}[t!]
		\begin{center}
			\includegraphics[width=0.45\textwidth]{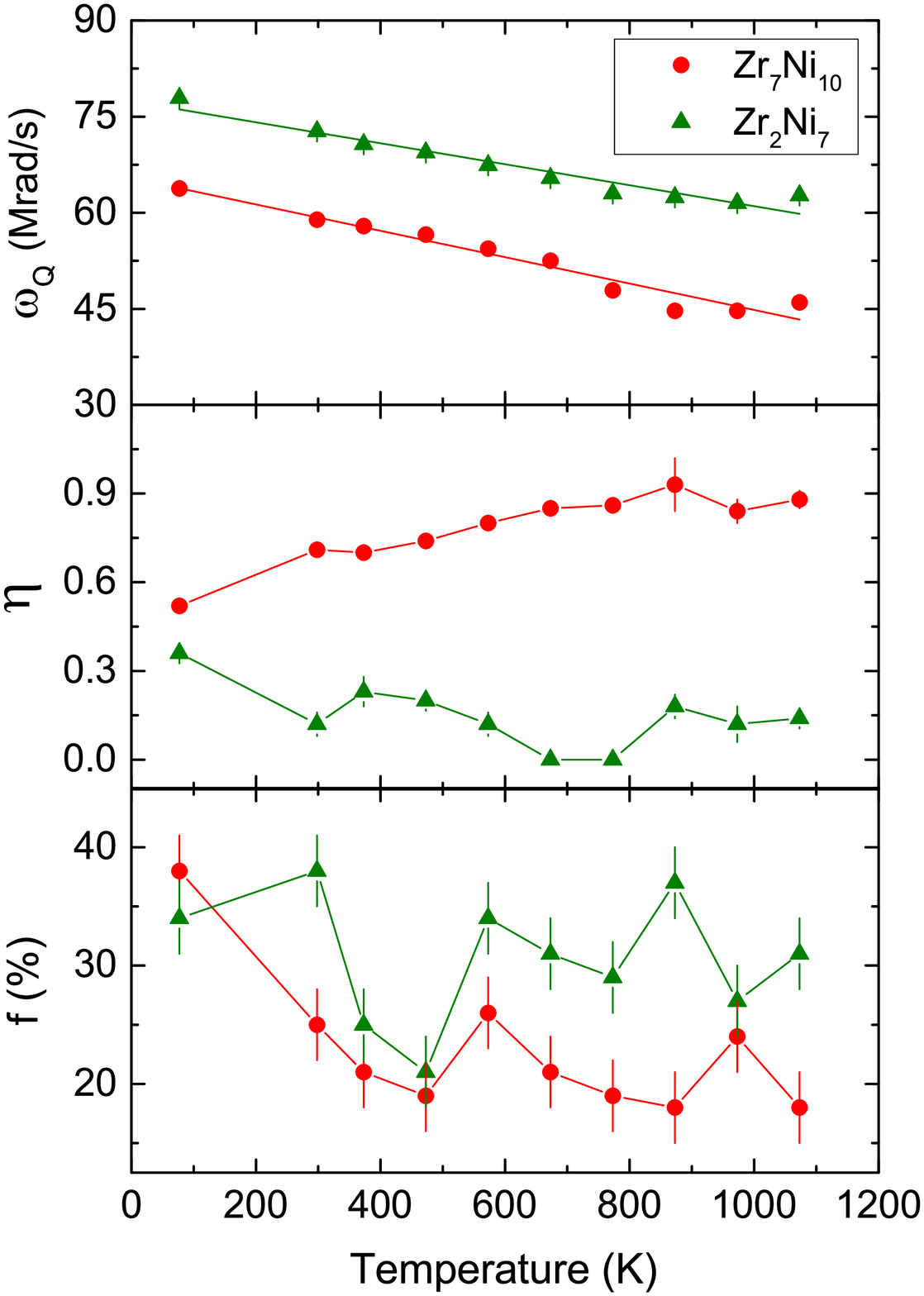}
		\end{center}
		\captionof{figure}{\small{Variations of quadrupole frequency ($\omega_Q$), asymmetry parameter ($\eta$) and site fraction $f$(\%)
				with temperature for the components of Zr$_7$Ni$_{10}$ and Zr$_2$Ni$_7$.}}
		\label{Zr7Ni10_parameter}
	\end{figure} 

\section{Results and discussion}
\subsection{Zr$_7$Ni$_{10}$}
The XRD powder pattern obtained in stoichiometric Zr$_7$Ni$_{10}$ sample is shown in Figure \ref{XRD_Zr7Ni10}. Peaks were first identified
using ICDD database. The presence of
orthorhombic Zr$_7$Ni$_{10}$ (\citep{J.-M. Joubert}, PDF Card No.: 01-072-3501), triclinic Zr$_8$Ni$_{21}$ (\citep{Krist}, PDF Card No.: 01-071-2622)
and monoclinic Zr$_2$Ni$_7$ (\citep{Eshelman}, PDF Card No.: 01-071-0543) phases have been found from XRD analysis. The x-ray intensity profile has been fitted using 
FullProf software package \citep{FullProf}. The presence of Zr$_7$Ni$_{10}$, Zr$_2$Ni$_7$ and Zr$_8$Ni$_{21}$ phases in
this stoichiometric sample of Zr$_7$Ni$_{10}$ have been observed from TEM/SAED measurement also (Figure \ref{TEM_Zr7Ni10}). The SAED pattern obtained from a region marked by a dotted circle in the stoichiometric sample of Zr$_7$Ni$_{10}$ is shown in Figure \ref{TEM_Zr7Ni10}. The interplaner spacing ($d_{hkl}$) is obtained by
measuring the distance ($\Delta q$) of a particular spot from the central bright spot using the formula $d_{hkl}=1/\Delta q$. Few of the 
measured $d_{hkl}$ from the SAED pattern
are 1.47(4) \r{A}, 2.26(4) \r{A} and 1.81(4) \r{A}. These measured interplaner spacings are very close to the (8 0 2), (4 2 2) and (4 4 1) inter-planer
spacings of orthorhombic Zr$_7$Ni$_{10}$ (ICDD PDF Card No.:01-072-3501), respectively. This further confirms the presence of Zr$_7$Ni$_{10}$ phase in the sample. 
Some of the measured $d_{hkl}$ from the SAED pattern
are  1.39(4) \r{A}, 1.49(4) \r{A}, 2.73(4) \r{A}, 3.39(4) \r{A}, 1.83(4) \r{A} and 1.36(4) \r{A} which are very close to the (-1 5 2), (1 5 2), (1 1 3), (0 2 2), (0 4 3) and (1 3 7) inter-planer
spacings of monoclinic Zr$_2$Ni$_7$ (JCPDS $\#$ 65-2321), respectively. The presence of Zr$_2$Ni$_7$ phase in the sample is thus confirmed. 
Few of the measured interplaner
spacings from the SAED pattern, 2.34(4) \r{A} and 3.38(4) \r{A}, are found to be very close to the (-1 2 2) and (1 2 1) inter-planer
spacing of triclinic Zr$_8$Ni$_{21}$ (ICDD PDF Card No.: 01-071-2622), respectively, which confirms the presence of Zr$_8$Ni$_{21}$ phase in the stoichiometric sample of Zr$_7$Ni$_{10}$.

The PAC spectrum in the stoichiometric Zr$_7$Ni$_{10}$ sample at room temperature is shown in Figure \ref{Zr7Ni10PAC}. 
The spectrum is found to be best fitted by considering five quadrupole frequencies. Texture effects are observed in the sample which indicates that
the EFGs produced at the probe sites are not randomly oriented and it is not a perfect polycrystalline sample. Therefore, the spectrum is analyzed using free $S_{2n}$ ($n$=0,1,2,3) parameters. At room temperature, a major component ($\sim$ 38\%) is found with values of $\omega_Q$ = 72.7(3) Mrad/s, $\eta$ = 0.12(4) and
$\delta$ = 1.1(7)\%. This component can be assigned to Zr$_2$Ni$_7$ by comparing with the results found in 
Zr$_2$Ni$_7$ \citep{Srivastava}. In the stoichiometric ZrNi$_3$ \citep{SKDEYHfNi3} and ZrNi$_5$ \citep{Dey} alloys also, the phase Zr$_2$Ni$_7$ was produced
as a major component.

 The second major component (Table \ref{tab:table_Zr7Ni10}) with values of $\omega_Q$ = 58.9(3) Mrad/s, $\eta$ = 0.71(1) is
attributed to Zr$_7$Ni$_{10}$ phase by comparing the result with our DFT calculation (discussed later). A minor component ($\sim$11\%) was found to be present with values of $\omega_Q$ = 77.1(9) Mrad/s, $\eta$ = 0.81(2). This component
has been identified as Zr$_8$Ni$_{21}$ by comparing the result with our recent PAC investigation
in Zr$_8$Ni$_{21}$ \citep{SKDeyZr8Ni21} for Zr(1) crystallographic site. Apart from these components, two other minor frequency components (Table \ref{tab:table_Zr7Ni10}) have been found. These two components are attributed to defects. Since the activation of the sample was done after preparing the sample, crystalline defects can be produced by neutron irradiation \citep{Dai}.  From the Zr-Ni phase diagram, it is found that the
phase Zr$_2$Ni$_7$ melts congruently and Zr$_8$Ni$_{21}$ phase is formed peritectically from Zr$_2$Ni$_7$ and liquid
melt (L+Zr$_2$Ni$_7\rightarrow$ Zr$_8$Ni$_{21}$) at 1453 K \citep{Nash_Zr_Ni}. The phases Zr$_7$Ni$_{10}$ and Zr$_8$Ni$_{21}$ are formed from liquid
alloy by an eutectic reaction (L$\rightarrow$Zr$_8$Ni$_{21}$+Zr$_7$Ni$_{10}$) \citep{Nash_Zr_Ni} at 1333 K. The phase Zr$_7$Ni$_{10}$ is also formed by
peritectic reaction
from Zr$_9$Ni$_{11}$ and liquid melt (L + Zr$_9$Ni$_{11} \rightarrow$ Zr$_7$Ni$_{10}$) at 1393 K \citep{Kosorukova}. 

At 77 K, the Zr$_2$Ni$_7$ and Zr$_7$Ni$_{10}$ phases were found only (Table \ref{tab:table_Zr7Ni10}). 
All the three Zr-Ni phases,
viz. Zr$_2$Ni$_7$, Zr$_8$Ni$_{21}$ and Zr$_7$Ni$_{10}$ are present in the
temperature range 298-773 K (Table \ref{tab:table_Zr7Ni10}). At 873 K and above, 
the component due to Zr$_8$Ni$_{21}$ does not appear. The phases Zr$_2$Ni$_7$ and Zr$_7$Ni$_{10}$, however, remain stable up to 1073 K.
The phase Zr$_2$Ni$_7$ is found to be predominant in the whole temperature range (77-1073 K) among the Zr-Ni binary phases that are produced in the
stoichiometric sample of Zr$_7$Ni$_{10}$. The frequency values for the fourth and fifth components show anomalous temperature dependence (Table \ref{tab:table_Zr7Ni10}). This further indicates that these components are irregular defect components. A re-measurement is carried at room temperature after measurement at 1073 K. Here, all the three
Zr-Ni phases, viz. Zr$_2$Ni$_7$, Zr$_7$Ni$_{10}$ and Zr$_8$Ni$_{21}$ produced with almost same fractions reversibly.

Temperature evolution
of quadrupole frequency, 
asymmetry parameter and site fraction for two Zr-Ni phases, viz. Zr$_2$Ni$_7$ and Zr$_{7}$Ni$_{10}$ present in the stoichiometric
sample of Zr$_7$Ni$_{10}$ are shown in Figure \ref{Zr7Ni10_parameter}. The asymmetry parameter of Zr$_7$Ni$_{10}$ phase is found to increase with temperature.
The quadrupole frequencies for both the components decrease linearly with temperature. 
The
values of quadrupole frequencies obtained
for Zr$_2$Ni$_7$ and Zr$_7$Ni$_{10}$ phases in the temperature range
77-1073 K have been fitted with the following relation
\begin{equation}
 \omega_Q(T) = \omega_Q(0) [1-\alpha T].
 \label{Eq:linearWQ}
\end{equation}
The fitted results give $\omega_Q$(0) = 65(1) Mrad/s ($V_{zz}$ = 7.3(2)$\times$10$^{21}$ V/m$^2$),
$\alpha$ = 3.1(3)$\times$10$^{-4}$ K$^{-1}$ for Zr$_7$Ni$_{10}$ component. For the Zr$_2$Ni$_7$ component, the results
are $\omega_Q$(0) = 77(1) Mrad/s ($V_{zz}$ = 8.6(2)$\times$10$^{21}$ V/m$^2$), $\alpha$ = 2.1(2)$\times$10$^{-4}$ K$^{-1}$. The linear temperature dependence 
of quadrupole frequency was observed in many intermetallic compounds \citep{Marszalek2,Wodniecka,Wodniecki,Kulinska,Kulinska2,Hrynkiewicz,Petrilli}. 

\begin{figure*}[t!]
\begin{center}
\includegraphics[width=0.8\textwidth]{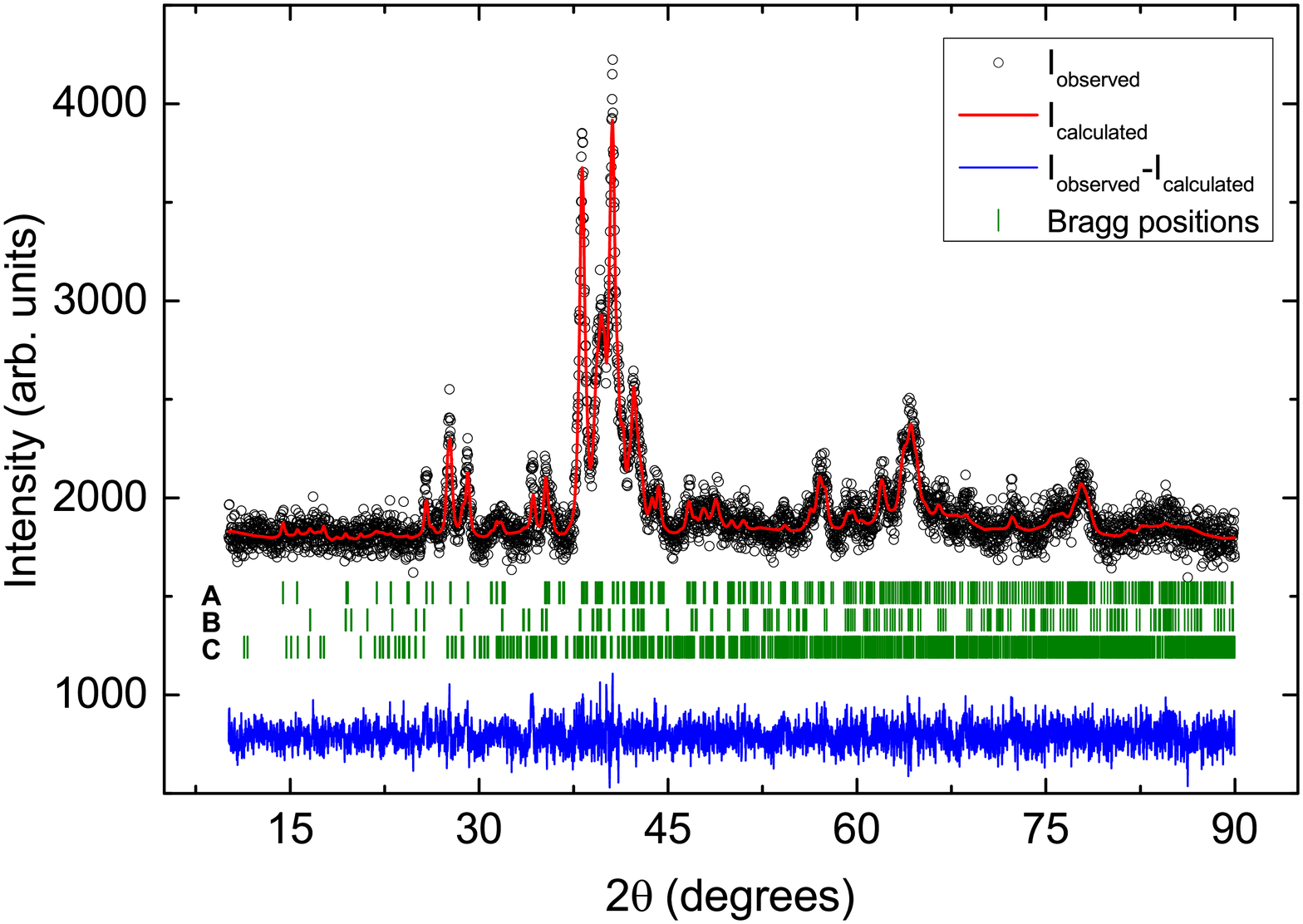}
\end{center}
\captionof{figure}{\small{The background subtracted XRD powder pattern in the stoichiometric sample of Hf$_7$Ni$_{10}$. 
The vertical bars A, B and C denote the Bragg angles corresponding to Hf$_7$Ni$_{10}$, $\beta$-HfNi$_3$ and Hf$_8$Ni$_{21}$, respectively.}}
\label{Hf7Ni10XRD}
\end{figure*} 
\begin{figure}
	\begin{center}
		\includegraphics[width=0.45\textwidth]{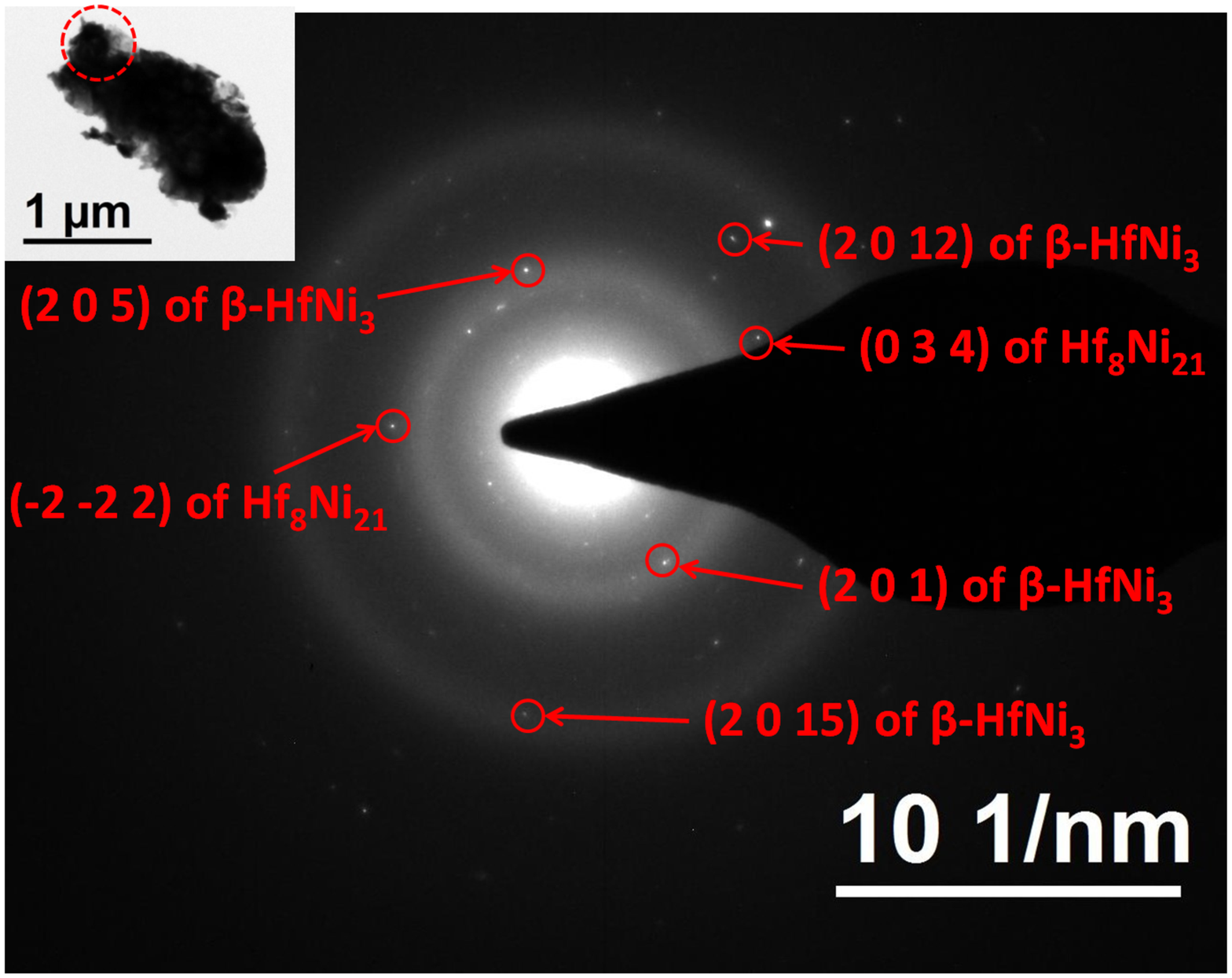}
	\end{center}
	\captionof{figure}{\small{Selected area electron diffraction pattern from stoichiometric
			Hf$_7$Ni$_{10}$ particle shown in the
			inset.}}
	\label{TEM_Hf7Ni10}
\end{figure}

\begin{figure*}
\begin{center}
\includegraphics[width=0.6\textwidth]{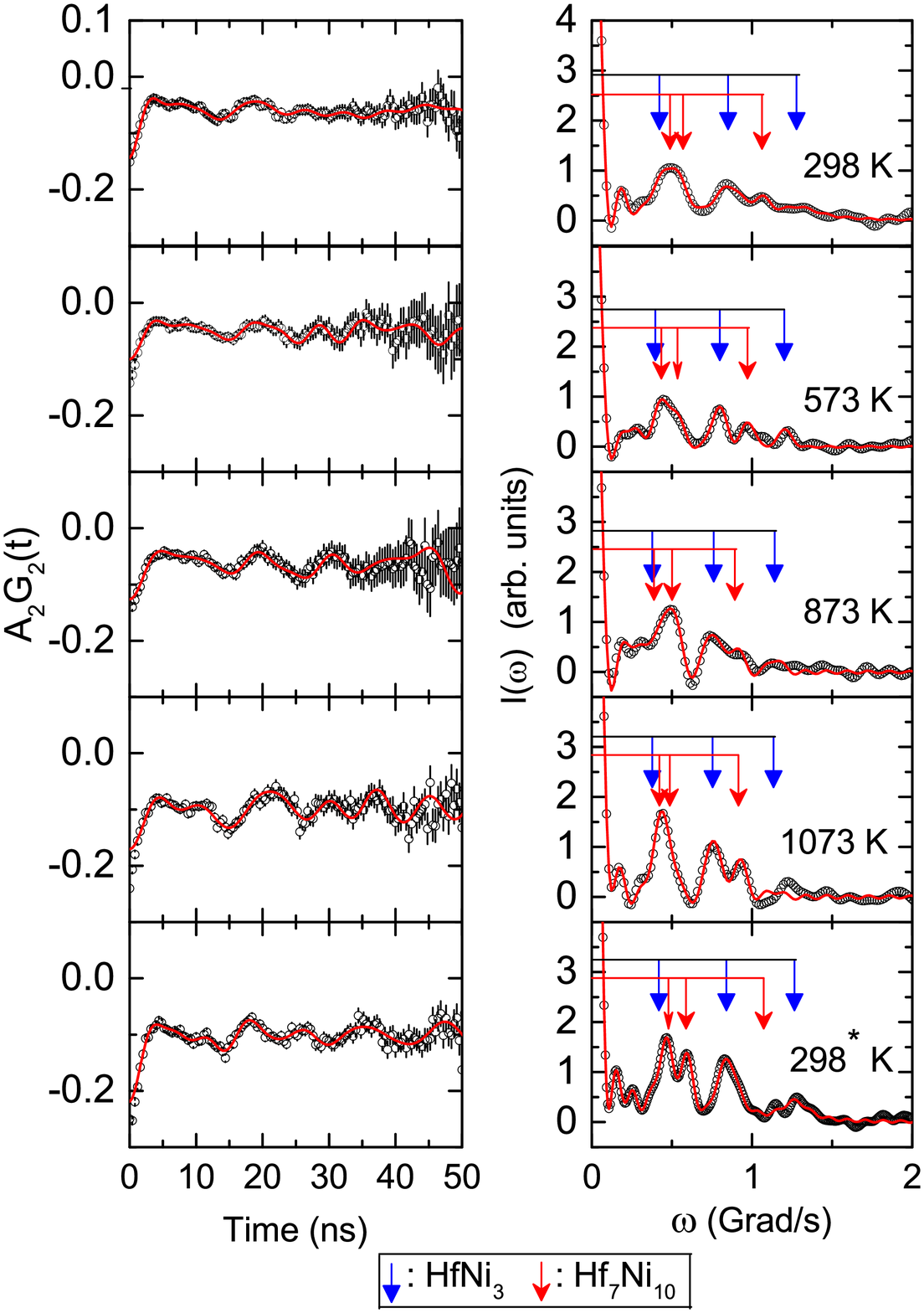}
\end{center}
\captionof{figure}{\small{Time differential perturbed angular correlation spectra in Hf$_7$Ni$_{10}$ at different temperatures. 
Left panel shows the time spectra and the right panel shows the corresponding Fourier cosine transforms. The PAC spectrum designated
by 298$^*$ K is taken at room
temperature after the measurement at 1073 K.}}
\label{Hf7Ni10PAC}
\end{figure*}

\begin{figure}[t!]
\begin{center}
\includegraphics[width=0.45\textwidth]{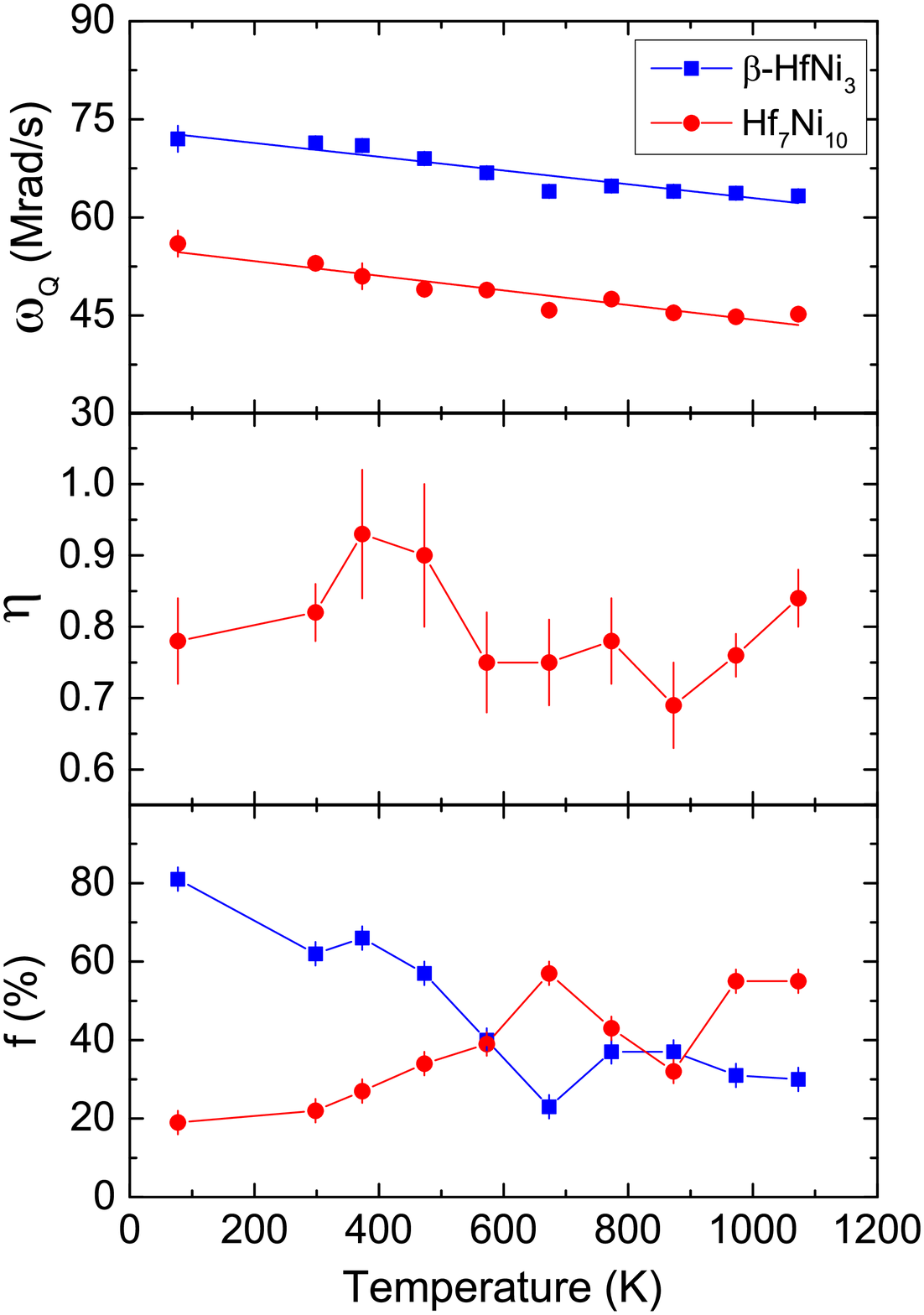}
\end{center}
\vspace{-0.5cm}
\captionof{figure}{\small{Variations of quadrupole frequency ($\omega_Q$), asymmetry parameter ($\eta$) and site fraction $f$(\%)
with temperature for the components of Hf$_7$Ni$_{10}$ and $\beta$-HfNi$_3$.}}
\label{Hf7Ni10_parameter}
\end{figure}
\subsection{Hf$_7$Ni$_{10}$}
The XRD powder pattern in the Hf$_7$Ni$_{10}$ sample is shown in Figure \ref{Hf7Ni10XRD}. The peaks were identified using ICDD database, 2009. Presence of the hexagonal $\beta$-HfNi$_3$ (\citep{HfNi3_crystal_structure}, PDF Card No.: 01-071-0475), orthorhombic Hf$_7$Ni$_{10}$ \citep{Kirkpatrick},
and triclinic Hf$_8$Ni$_{21}$ (\citep{Hf8Ni21_crystal_structure}, PDF Card No.: 01-071-0476) phases
are found in the stoichiometric Hf$_7$Ni$_{10}$ sample. Fitting of the XRD intensity profile has been carried out
using FullProf software package \citep{FullProf}. The space group for Hf$_7$Ni$_{10}$ has been considered $Cmca$ due to its 
isostructurality with Zr$_7$Ni$_{10}$ \citep{Kirkpatrick, PAnash_Hf_Ni}. Selected area electron
diffraction (SAED) pattern obtained from a region marked by a dotted circle in the stoichiometric sample of Hf$_7$Ni$_{10}$ is shown in Fig. \ref{TEM_Hf7Ni10}. Some of the 
measured interplaner
spacing from the SAED pattern
are 1.41(4) \r{A}, 2.27(4) \r{A} and 2.01(4) \r{A} which are found to be very close to the (2 0 12), (2 0 1) and (2 0 5) inter-planer
spacings of hexagonal $\beta$-HfNi$_3$ (JCPDS $\#$71-0475). This further confirms the presence of HfNi$_3$ phase in the sample. Few of the interplaner spacing from the SAED pattern are 1.70(4) \r{A} and 1.82(4) \r{A}. These are found to be very close to the (0 3 4) and (-2 -2 2) interplaner spacings of triclinic Hf$_8$Ni$_{21}$ (JCPDS $\#$71-0476). The phase Hf$_7$Ni$_{10}$ 
could not be identified from SAED pattern in the stoichiometric sample of Hf$_7$Ni$_{10}$ due to non-availability of x-ray diffraction data of interplaner spacings ($d_{hkl}$) and corresponding crystallographic planes ($h\;k\;l$) for Hf$_7$Ni$_{10}$.

The PAC spectrum at room temperature in the stoichiometric Hf$_7$Ni$_{10}$ sample is shown in Figure \ref{Hf7Ni10PAC}. Fitting of the spectrum shows the presence of three quadrupole 
frequency components. Analysis of the spectrum has been carried out using free $S_{2n}$ ($n$=0,1,2,3) parameters as the sample is found to have texture effects. 
The predominant component ($\sim$ 62\%) produces values of
$\omega_Q$ = 71.4(9) Mrad/s, $\eta$ = 0, $\delta$ = 11(2)\%. This component has been assigned
to $\beta$-HfNi$_3$ by comparing the values with the results from
our recent PAC investigation in HfNi$_3$ sample \citep{SKDEYHfNi3} for Hf(3) crystallographic site. The second component produces values
of $\omega_Q$ = 53(1) Mrad/s, $\eta$ = 0.82(4) with site fraction $\sim$ 22\%. This component has been assigned to Hf$_7$Ni$_{10}$ by comparing
the values with the results found in the analogous Zr$_7$Ni$_{10}$ phase (Table \ref{tab:table_Zr7Ni10}) and with the results from
DFT calculations (discussed later). A minor component ($\sim$16\%) with values of $\omega_Q$ = 32(1) Mrad/s, $\eta$ = 0 was also
found to be present. The value of quadrupole frequency for this component varies with temperature in an irregular manner and suggests that this is a crystalline defect produced during sample preparation. In the PAC sample, however, no component due to Hf$_8$Ni$_{21}$ was observed. In the Hf-Ni 
phase diagram, it is found that Hf$_2$Ni$_7$ and HfNi are congruently melting phases. The HfNi$_3$ has two polymorphs, one is high temperature phase ($\alpha$-HfNi$_3$) 
which is stable above 1473 K and the other is low temperature phase ($\beta$-HfNi$_3$) which is stable below 1473 K \citep{Bsenko}. The phase HfNi$_3$ is
produced by peritectic reaction of liquid melt with Hf$_2$Ni$_7$ (L+Hf$_2$Ni$_7\rightarrow$HfNi$_3$) at 1573 K. It was
reported \citep{Kejun} that the phase Hf$_7$Ni$_{10}$ is produced
by peritectic reaction L + Hf$_9$Ni$_{11} \rightarrow$ Hf$_7$Ni$_{10}$ \citep{Kejun} at $\sim$1563 K.    
\begin{table}[t!]
 \begin{center}
\captionof{table}{\small{ Results of PAC measurements in Hf$_7$Ni$_{10}$}}
\label{tab:table_Hf7Ni10} 
\scalebox{0.65}{
\begin{tabular}{l l l l l l l l l l l} 
 \hline  \\ [-0.9ex]
Temperature (K)   &Component     & $\omega_Q$ (Mrad/s)     & $\eta$     & $\delta$($\%$)   & $f$($\%$)      & Specification     \\ [1.5ex]
 \hline  
77                & 1    & 72(2)                 & 0          & 11(3)                & 81(3)                & $\beta$-HfNi$_3$      \\
                   &2    & 56(2)                 & 0.78(6)          & 0               & 19(3)          &   Hf$_7$Ni$_{10}$  \\   \hline                        
298                &1    & 71.4(9)                 & 0          & 11(2)                & 62(3)                 & $\beta$-HfNi$_3$     \\
                   &2    & 53(1)                & 0.82(4)          & 0                & 22(3)         & Hf$_7$Ni$_{10}$      \\    
                   &3    & 32(1)                 & 0          & 0                & 16(3)             &  \\ \hline
373                &1    & 71(1)                & 0          & 7(2)                & 66(3)    & $\beta$-HfNi$_3$        \\  
                   &2   & 51(2)                 & 0.93(9)          & 0                & 27(3)    &  Hf$_7$Ni$_{10}$         \\   
                   &3    & 38(4)                 & 0         & 0                & 7(3)     &            \\ \hline 
473                &1    & 69(1)               & 0          & 8(3)                & 57(3)       & $\beta$-HfNi$_3$      \\  
                   &2    & 49(1)                 & 0.9(1)          & 0                & 34(3)             &   Hf$_7$Ni$_{10}$ \\   
                    &3   & 33(3)                 & 0          & 0                & 9(3)               & \\ \hline 
573                &1    & 66.8(6)                 & 0          & 0                & 40(3)              & $\beta$-HfNi$_3$\\ 
		    &2   & 48.9(9)                & 0.75(7)          & 0                & 39(3)       &  Hf$_7$Ni$_{10}$    \\    
                    &3   & 42(2)                 & 0          & 0                & 21(3)               &  \\ \hline  
673                &1    & 64.0(8)                & 0          & 0                & 23(3)     &  $\beta$-HfNi$_3$        \\   
                   &2    & 45.8(6)                 & 0.75(6)          & 0                & 57(3)              & Hf$_7$Ni$_{10}$\\ 
                   &3    & 45(2)                 & 0          & 0                & 20(3)           &      \\ \hline 
773                &1    & 64.8(7)                & 0          & 0                & 37(3)     & $\beta$-HfNi$_3$    \\   
                   &2    & 47.5(7)                 & 0.78(6)          & 0                & 43(3)              & Hf$_7$Ni$_{10}$ \\ 
                   &3    & 39(2)                 & 0          & 0                & 20(3)           &     \\ \hline     
873                &1    & 64.0(1)                & 0          & 0                & 37(3)     &  $\beta$-HfNi$_3$      \\   
                   &2    & 45.4(6)                 & 0.69(6)          & 0                & 32(3)              & Hf$_7$Ni$_{10}$ \\ 
                   &3    & 40(2)                 & 0          & 0                & 30(3)           &      \\ \hline     
973                &1    & 63.7(4)                & 0          & 0                & 31(3)     &  $\beta$-HfNi$_3$     \\   
                   &2    & 44.8(4)                 & 0.76(3)          & 0                & 55(3)              & Hf$_7$Ni$_{10}$ \\ 
                   &3    & 16(3)                 & 0          & 0                & 14(3)           &     \\ \hline  
1073                &1    & 63.3(4)                & 0          & 0                & 30(3)     &  $\beta$-HfNi$_3$    \\   
                   &2    & 45.2(5)                 & 0.84(4)          & 0                & 55(3)              & Hf$_7$Ni$_{10}$ \\ 
                   &3    & 10(1)                 & 0          & 0                & 15(3)           &      \\ \hline
298$^*$           &1     & 70.5(4)                 & 0          & 7(1)                & 58(3)   & $\beta$-HfNi$_3$       \\   
                   &2    & 53.9(4)                 & 0.76(1)          & 0                & 25(3)             & Hf$_7$Ni$_{10}$   \\ 
                   &3    & 15.9(9)                 & 0          & 0                & 16(3)             &  \\                                                                                     
 \hline
\end{tabular}}
\begin{center}
 \scriptsize{$^*$ After measurement at 1073 K } 
\end{center}
\end{center}
\end{table}
At 77 K, the phase fraction of $\beta$-HfNi$_3$ component
enhances to $\sim$ 81\%. The site fraction of Hf$_7$Ni$_{10}$ increases while the site percentage of $\beta$-HfNi$_3$ phase
decreases up to 673 K (Figure \ref{Hf7Ni10_parameter}) with the increase of temperature. However, the phase $\beta$-HfNi$_3$ was found
to be predominant up to 573 K (Table \ref{tab:table_Hf7Ni10}). The Hf$_7$Ni$_{10}$ phase becomes predominant ($\sim$57\%) and
the phase fraction of $\beta$-HfNi$_3$ decrease to 23\% at 673 K. In the temperature range 773-873 K, the phase fraction of Hf$_7$Ni$_{10}$ decreases
and $\beta$-HfNi$_3$ increases to 37\%. Major contribution ($\sim$55\%) in the PAC spectrum at 973 K and 1073 K was found due to the Hf$_7$Ni$_{10}$ phase and $\beta$-HfNi$_3$ phase
fraction reduces to $\sim$30\%. The PAC measurement was carried out at room temperature after measurement at 1073 K. Similar results of phase fractions of $\beta$-HfNi$_3$ and Hf$_7$Ni$_{10}$ at initial room temperature and after 1073 K (Table \ref{tab:table_Hf7Ni10}) show phase reversibility of the two phases.

Variation of quadrupole frequency ($\omega_Q$), asymmetry parameter ($\eta$) and phase fraction with temperature for $\beta$-HfNi$_3$ and Hf$_7$Ni$_{10}$ phases are shown in 
Figure \ref{Hf7Ni10_parameter}. 
The quadrupole frequency for $\beta$-HfNi$_3$ and Hf$_7$Ni$_{10}$ phases decrease linearly with temperature following the Eqn. \ref{Eq:linearWQ}. Similar temperature dependence of $\omega_Q$ was found in Zr$_7$Ni$_{10}$ which shows isostructurality of Zr$_7$Ni$_{10}$ and Hf$_7$Ni$_{10}$. The fitted results are
$\omega_Q$(0) = 55.9(9) Mrad/s ($V_{zz}$=6.2(2)$\times$10$^{21}$ V/m$^2$), $\alpha$ = 2.0(2)$\times$10$^{-4}$ K$^{-1}$ for the Hf$_7$Ni$_{10}$ phase. For
$\beta$-HfNi$_3$ phase, the fitted results are $\omega_Q$(0) = 73.5(9) Mrad/s ($V_{zz}$=8.2(2)$\times$10$^{21}$ V/m$^2$), $\alpha$ = 1.5(1)$\times$10$^{-4}$ K$^{-1}$.
\section{DFT calculations and results}
Zr$_7$Ni$_{10}$ and Hf$_7$Ni$_{10}$  crystallize in the orthorhombic base-centered centrosymmetric $Cmca$ type structure (space group number 64) with lattice parameters $a$=12.381 \r{A}, $b$=9.185 \r{A} and $c$=9.221 \r{A} for Zr$_7$Ni$_{10}$ \citep{J.-M. Joubert} and $a$=12.275 \r{A}, $b$=9.078 \r{A}, $c$=9.126 \r{A} for Hf$_7$Ni$_{10}$ \citep{PAnash_Hf_Ni}. This structure contains 34 atoms in the unit cell and possesses 7 non-equivalent crystallographic positions; 4 non-equivalent positions for Zr (or Hf) and 3 non-equivalent positions for Ni.
	
First we have optimized these structural parameters. The first-principles density functional theory (DFT) calculations were performed to compare with the experimental results. All the calculations were done using the full potential (linearized) augmented plane waves method [FP-(L)APW], as implemented in WIEN2k \citep{PBlaha}. The energy convergence has been achieved by expanding the basis function up to $R_{MT}\cdot K_{max}$=7, where $R_{MT}$ is the smallest atomic sphere radius in the unit cell and $K_{max}$ gives the magnitude of the largest $\vec{k}$ vector in the plane wave expansion. In our calculations the muffin-tin radii for Zr, Ni and Hf(Ta) were 2.2, 2.1 and 2.15 a. u., respectively. The valence wave functions inside the spheres are expanded up to $l_{max}$=10 while the charge density is Fourier expanded up to $G_{max}$=16. The energy to separate core and valence states was set to -7 Ry. Electronic exchange-correlation energy was treated with generalized gradient approximation (GGA) parametrized by Perdew-Burke-Ernzerhof (PBE) \citep{Perdew}. Taking into consideration both the accuracy and the efficiency of the calculations, we have selected a 8$\times$8$\times$8 $k$ point mesh to sample the entire Brillouin-zone (BZ), yielding 143 points in the irreducible Brillouin-zone. The structure was relaxed according to Hellmann-Feynman forces calculated at the end of each self-consistent cycle, until the forces acting on all atoms were less than 0.068 eV/\r{A} (5 mRy/a.u.). The relaxation method is described in Ref. \citep{Koteski}. In our calculations the self-consistency was achieved by demanding the convergence of the integrated charge difference between last two iterations to be smaller than 10$^{-5}$e. All the calculations refere to zero temperature.

The theoretically optimized lattice parameters $a$, $b$ and $c$, and fractional coordinates of atoms together with the present and previous experimental values \citep{J.-M. Joubert,PAnash_Hf_Ni} are presented in Table \ref{tab:Zr7Ni10_Hf7Ni10_structure}. From Table \ref{tab:Zr7Ni10_Hf7Ni10_structure} it can be seen that our calculated parameters are in very good agreement with the experimental results.
After obtaining the optimized structural parameters, we replaced one of the host sites; i.e. one of the 4 non-equivalent positions of Zr (or Hf) by a Ta atom (preserving the point group symmetry around original atom), in order to simulate a dopant in the crystal lattice. This substitutional structures have been marked as X1-Ta, X2-Ta, X3-Ta and X4-Ta; X=Zr, Hf. For each case of the substitutional structure, we have repeated calculations again, keeping all parameters and charge convergence criteria same as in the case of the pure compounds. For example, to simulate PAC measurements at Zr1 position, we replaced one  atom at position (0, 0.31359, 0.18707) with Ta atom. We checked that the two Ta atoms are sufficiently far from each other ($\sim$8 \r{A}) to avoid significant impurity-impurity interactions.
The calculated electric field gradients (EFGs) in the pure compounds as well as at Ta probe positions in the Zr$_7$Ni$_{10}$ and Hf$_7$Ni$_{10}$ structures along with the values of asymmetry parameter $\eta$, are given in Table \ref{tab:DFT_Zr7Ni10_Hf7Ni10}. The sign of EFG ($V_{zz}$) can not be determined from PAC measurement. Thus, absolute values of measured EFG (extrapolated to 0 K) and asymmetry parameter (at 77 K) for Zr$_7$Ni$_{10}$ and Hf$_7$Ni$_{10}$ have been compared with the theoretical results in the Table \ref{tab:DFT_Zr7Ni10_Hf7Ni10}. The calculation of EFG were performed by using the method developed in Ref. \citep{Herzig}; which is implemented in WIEN2k code. All the calculations refer to zero temperature.

We see that the calculated result for EFG at the Ta probe site replacing Zr3 atom
(6.99$\times$10$^{21}$ V/m$^2$) with asymmetry parameter 0.54 is in excellent agreement with the measured value of EFG extrapolated to 0 K (7.3(2)$\times$10$^{21}$ V/m$^2$) and $\eta$ (77 K)=0.52(1) for the component Zr$_7$Ni$_{10}$, thus confirming that the mentioned component of the measured PAC spectra originates from Zr$_7$Ni$_{10}$. Similarly, the calculated
result for EFG at the Ta probe site replacing Hf3 atom (6.37$\times$10$^{21}$ V/m$^2$ ) with asymmetry parameter 0.77 is in excellent agreement with the measured value of EFG extrapolated to 0 K (6.2(2)$\times$10$^{21}$ V/m$^2$) and $\eta$ (77 K)=0.78(6) for the component Hf$_7$Ni$_{10}$, thus confirming that the mentioned component of the measured PAC spectra originates from Hf$_7$Ni$_{10}$.

\begin{table}[t!]
	\begin{center} 
		\captionof{table}{\small{The lattice constants a, b and c (given in \r{A}) of the Zr$_7$Ni$_{10}$ and Hf$_7$Ni$_{10}$ $Cmca$ crystal
				structure and the fractional coordinates of 7 crystallographic non-equivalent positions.}}
		\scalebox{0.67}{
			\begin{tabular}{ l l l l} 
				\hline  \\ [-0.9ex]
				&Present calculated results   & Experimental results \citep{J.-M. Joubert}, \citep{PAnash_Hf_Ni}  &Present experimental  \\ 
				&(WIEN2k)&(XRD)& results (XRD)\\ [1.5ex]
				\hline  \\

				Zr$_7$Ni$_{10}$ & & \\	
				$a$ &12.365 &12.381 &12.374\\
				$b$ &9.172 &9.185 &9.173\\
				$c$ &9.197 &9.221&9.213\\
				Zr1  &0\; 0.31359 \;0.18707 &0\; 0.31219 \;0.18847  \\
				Zr2  &1/4 \; 0.25466 \;1/4 & 1/4 \;0.25466\; 1/4  \\
				Zr3 &0.30754 \;0 \;0 &0.30634 \;0\; 0  \\
				Zr4  &0\; 0 \;0 & 0\; 0\; 0  \\
				Ni1  &0.14438\; 0.01115 \;0.20822 &0.14438\; 0.01115\; 0.20822 \\
				Ni2  &0.35507 \;0.29157 \;0.00833 & 0.35507 \;0.29157 \;0.00833  \\
				Ni3  &0\; 0.10655 \;0.39553 &0\; 0.10755 \;0.39423 \\ \\
				
				Hf$_7$Ni$_{10}$ & & \\
				$a$ &12.281 &12.275&12.279 \\
				$b$ &9.062 &9.078&9.071 \\
				$c$ &9.151 &9.126&9.120\\
				Hf1 &0\; 0.31439 \;0.1867 &0\; 0.31219 \;0.18847  \\
				Hf2  &1/4\; 0.25504 \;1/4 &1/4\; 0.25466 \;1/4  \\
				Hf3  &0.30645 \;0 \;0 &0.30634 \;0 \;0  \\
				Hf4  &0\; 0 \;0 &0\; 0 \;0  \\
				Ni1  &0.14391 \;0.00998 \;0.20874 &0.14438 \;0.01115 \;0.20822  \\
				Ni2  &0.35507 \;0.29111 \;0.00665 &0.35507 \;0.29157 \;0.00833 \\
				Ni3  &0 \;0.10699 \;0.39569 &0 \;0.10755 \;0.39423 \\ \\ 
				\hline
			\end{tabular}}
			\label{tab:Zr7Ni10_Hf7Ni10_structure}
		\end{center}
	\end{table}

\begin{table}[t!]
	\begin{center} 
		\captionof{table}{\small{  The calculated and experimental EFG values in units of 10$^{21}$ V/m$^2$ and asymmetry parameter ($\eta$) values for Zr$_7$Ni$_{10}$ and Hf$_7$Ni$_{10}$	$Cmca$ crystal structure.}}
		\scalebox{0.75}{
			\begin{tabular}{ l l l l l l} 
				\hline  \\ [-0.9ex]
				Probe  &Lattice Site    & calculated  & calculated  & Measured & Measured \\ 
				& site & EFG &$\eta$& EFG & $\eta$ (77 K)    \\
				\hline \\ 
				Zr$_7$Ni$_{10}$ & & & &  & \\
	no probe    &Zr1 &1.11 &0.63 & & \\
(pure compound) &Zr2  &4.05 &0.10 & &\\
				&Zr3  &-3.02 &0.37 & &\\ 
				&Zr4   &-3.38 &0.41 & &\\ \\
	$^{181}$Ta    &Zr1-Ta &1.33 &0.77 & & \\
	 &Zr2-Ta  &10.11 &0.10 & & \\
	&Zr3-Ta  &-6.99 &0.54 &7.3(2) &0.52(1)\\ 
	&Zr4-Ta  &-10.75 &0.25 \\ \\	
	\hline			\\
				Hf$_7$Ni$_{10}$ & & & & & \\
				no probe    &Hf1 &-1.15 &0.33 & & \\
				(pure compound) &Hf2  &9.48 &0.18 & & \\
				&Hf3  &-4.96 &0.82 & &\\ 
				&Hf4   &-8.94 &0.32  & &\\ \\
				$^{181}$Ta    &Hf1-Ta &1.87 &0.25 & & \\
				&Hf2-Ta  &9.75 &0.10  & &\\
				&Hf3-Ta  &-6.37 &0.77  &6.2(2)&0.78(6)\\ 
				&Hf4-Ta  &-10.68 &0.22 \\ \\	
				\hline			\end{tabular}}
			\label{tab:DFT_Zr7Ni10_Hf7Ni10}
		\end{center}
	\end{table}

\section{Conclusion}
 In stoichiometric Zr$_7$Ni$_{10}$ sample, the phases Zr$_2$Ni$_7$, Zr$_7$Ni$_{10}$ and Zr$_8$Ni$_{21}$ are produced where
Zr$_2$Ni$_7$ is found as a major phase and a minor phase due to Zr$_8$Ni$_{21}$ is found at room temperature. In the stoichiometric Hf$_7$Ni$_{10}$ sample, 
the phases $\beta$-HfNi$_3$ and Hf$_7$Ni$_{10}$ are produced where the phase $\beta$-HfNi$_3$ is predominant at room temperature. The phase fraction of Hf$_7$Ni$_{10}$ increases with temperature at the expense of  
$\beta$-HfNi$_3$. At  temperatures $\ge$ 400$^\circ$C, the phase Hf$_7$Ni$_{10}$ becomes predominant which indicates that it is a high temperature phase. However, these phase fractions are found to be reversible with temperature.  Similar 
values of quadrupole frequency and asymmetry parameter indicate isostructurality of Zr$_7$Ni$_{10}$ and Hf$_7$Ni$_{10}$ phases. In both Zr$_7$Ni$_{10}$ and Hf$_7$Ni$_{10}$, four non-equivalent crystallographic sites of Zr/Hf  have been found. Our experimental results of EFG and $\eta$ are in excellent agreement with the values of EFG at $^{181}$Ta sites corresponding to Zr3/Hf3 positions calculated by the first-principles density functional theory based on
the full potential (linearized) augmented plane waves method
[FP-(L)APW]. The origin of observed EFG in these materials can thus be explained. 


  \vspace{1 cm}
 {\bf Acknowledgement\\}

 The authors thankfully acknowledge Mr. A. Karmahapatra of Saha Institute of Nuclear Physics, Kolkata for X-ray diffraction measurements. The present work is
 supported by the Department of Atomic Energy, Government of
 India through the Grant no. 12-R$\&$D-SIN-5.02-0102 and by The Ministry of Education, Science and Technological Department of the Republic of
 Serbia through the Grant no. 171001.

\end{document}